\documentstyle[12pt]{article}

\begin{document}

\advance\textwidth 2cm
\advance\oddsidemargin -1cm
\title{Quantum Backreaction\\ on ``Classical" Variables}

\author{Arlen Anderson\thanks{arley@physics.unc.edu}
\thanks{Present address:
Dept. Physics, UNC-Chapel Hill, Chapel Hill NC 27599-3255.}\\
Isaac Newton Institute\\
20 Clarkson Road\\
Cambridge CB3 0EH, England\\
and\\
Blackett Laboratory\\
Imperial College\\
Prince Consort Rd.\\
London SW7 2BZ, England}
\date{June 26, 1994}

\maketitle
\vspace{-12cm}
\hfill Imperial-TP-93-94-44

\hfill hep-th/9406182
\vspace{11cm}

\begin{abstract}
A mathematically consistent procedure for coupling quasiclassical
and quantum variables through coupled Hamilton-Heisenberg equations of motion
is derived from a variational principle.  During evolution, the quasiclassical
variables become entangled with the quantum variables with the result
that the value of the quasiclassical variables depends on the quantum
state.  This provides a formalism to compute the backreaction of any
quantum system on a quasiclassical one. In particular, it leads to
a natural candidate for a theory of gravity coupled to quantized matter
in which the gravitational field is not quantized.
\end{abstract}
\newpage

Few would argue with the view that the universe is wholly quantum, and that
classical physics is only an approximate description of a collection of
phenomena that emerge in certain situations\cite{Har,GMH}. This view has
not always been universally held, especially concerning the quantization of
the gravitational field\cite{Ros}.
In this Letter, working within the simplified context of quantum mechanics,
I shall present and interpret a mathematically consistent scheme for
coupling ``classical'' and quantum variables. This has a dual purpose. It
is intended to assuage mathematical doubts about how classical and quantum
variables may coexist in a single theory. Secondly,
this approach serves as an approximate description of any
fully quantum mechanical theory when some of the variables behave
essentially classically. It provides a means of computing the backreaction
that quantum variables have on the evolution of classical ones without
having to make a full semiclassical analysis.

Confidently assuming that the same scheme extends to field theory, this
approach has potentially far-reaching implications. By presenting the
outline of a new theory in which the gravitational field is not quantized,
it reopens the debate on the necessity of quantizing gravity\cite{Ros,Kib}.
Alternatively, when viewed as an approximation to a fully quantum theory of
gravity in energy regimes where quantum gravitational effects can be
ignored, it improves on conventional quantum field theory in curved
spacetime by taking account of the backreaction of quantum matter fields on
the classical spacetime background. The effects of backreaction are
expected to be very important for a complete understanding of information
loss in black hole evaporation\cite{Page}. It is an interesting question
whether the predictions of a fully quantum theory of gravity would differ
in an experimentally detectable way from those of this type of formulation.

The approach followed here is similar in motivation to one discussed
independently by Aleksandrov\cite{Ale} and by Boucher and
Traschen\cite{BoT}, but it differs in details and has an
entirely original derivation and interpretation. The key feature is that
the classical variables evolve to become correlated with the state of the
quantum variables. Because this correlation may be with different states in a
quantum superposition, the classical variables need not have a definite
value but may take a distribution of values depending on the quantum state.
This parallels the behavior one expects in a fully quantum system as some
variables become classical (cf. \cite{GMH}). These variables are not then
entirely classical, but may be more properly called {\it quasiclassical}.

This behavior stands in contrast to that predicted by the traditional
approach of coupling classical variables to the expectation value of
quantum ones\cite{Ros,Kib}. As emphasized by Kibble\cite{Kib}, the
traditional approach amounts to a nonlinear modification of quantum theory.
The quasiclassical approach takes the easier route of accepting linear
quantum mechanics yet allowing the value of the quasiclassical variables to
depend on the quantum state. By relaxing the requirement that
quasiclassical variables take a single definite value, a
state-dependent backreaction of the quantum variables on the quasiclassical
evolution is made possible.   Effectively the quantum variables act like
a source of noise on the quasiclassical evolution.  (In this
connection, the approach of
Hu\cite{Hu} to include stochastic corrections to traditional semiclassical
gravity should be mentioned as a possibly related alternative.) One might
ask in what sense the quasiclassical variables are classical if it is not
because they take definite values. The answer is that they are classical
because they do not exhibit self-interference effects. That is, in the
absence of coupling to a quantum system, the quasiclassical variables
evolve classically without interference phenomena.

To elaborate on this, consider a fully quantum system in which some variables
evolve essentially classically, that is, as well localized wave packets
with minimal dispersion, when coupling to the remaining variables is
neglected.  In the presence of coupling,  quantum fluctuations of these other
variables can lead the ``classical evolution'' to divide into
a collection of qualitatively distinct evolutions,
most easily by triggering instabilities.  By appealing
to the concept of decoherence\cite{Har,GMH,Zur}, one finds
that each of these evolutions occurs with a classical probability
as the overlap between states associated to different
evolutions tends to zero.  This phenomenon in which the distribution
among possible quasiclassical evolutions is controlled by
quantum fluctuations in the non-quasiclassical variables is
potentially important in many areas of physics where the
classical and quantum regimes overlap.
It may be particularly important in the early universe. In
a different context, quantum fluctuations of a scalar field amplified by
inflation have already been proposed as the source of fluctuations in the
cosmic microwave background radiation and as seeds for galaxy
formation\cite{cmbr}.

The first step in the proposal for coupling quasiclassical and quantum
variables is to characterize each type of variable.  Viewed algebraically,
quantum variables are noncommutative variables satisfying commutation
relations. To be explicit, consider one variable and its conjugate with
the canonical commutation relation $[q,p]=i$ ($\hbar=1$). One can define a
noncommutative algebra ${\cal U}^{(q)}$ of functions of $q,\ p$ involving
arbitrary complex powers of each. This is essentially the algebra of
pseudodifferential operators, and it contains all observables, as well as
non-self-adjoint elements. Quantum canonical transformations are maps
between canonically conjugate pairs of elements which preserve the
canonical commutation relations\cite{qct,HBJ}. They are generated by the
adjoint action of an element, say, $C\in {\cal U}^{(q)}$, $C:(q,p)\mapsto
(CqC^{-1},CpC^{-1})$.

The Hamiltonian $H(q,p,t)$ at a fixed time is an element of ${\cal
U}^{(q)}$, and it can be used to define a quantum action,
\begin{equation}
\label{qaction}
S=\int_{t_1}^{t_2} dt\, {1\over 2} (p\dot q +\dot q p)-H(q,p,t),
\end{equation}
where the integral is over a path in ${\cal U}^{(q)}$. The Heisenberg
equations of motion can be derived by a quantum variational principle.
Variations are defined in terms of the infinitesimal canonical
transformations which generate translations of $q$ and $p$. Requiring that
the action be stationary under such variations with fixed endpoints leads
to the Heisenberg equations of motion\cite{ncact}. A canonical
transformation is simply a change of basis in ${\cal U}^{(q)}$, and
the Heisenberg equations of motion are invariant under them. For a
time-independent Hamiltonian, evolution in the Heisenberg picture is
produced by the unitary canonical transformation $C=e^{iHt}$.

(Quasi)classical variables satisfy Poisson bracket relations. Again,
consider the case of one variable with $\{x,k\}=1$ ($\{f,g\}=\partial_x
f \partial_k g- \partial_k f \partial_x g$). Classical mechanics can be
formulated by working from a
commutative algebra of functions ${\cal U}^{(c)}$ of the phase space
variables. Canonical transformations are maps between canonically conjugate
pairs of elements of this algebra. The action can be defined as an integral
over a path in ${\cal U}^{(c)}$ as in (\ref{qaction}). In parallel to the
quantum case, the Hamilton equations of motion are obtained by requiring
that the action is stationary under variations generated by the
infinitesimal canonical transformations which generate translations in $x$
and $k$. For a time-independent Hamiltonian, evolution is produced by the
unitary transformation $C=e^{v_H t}$, where $v_H f=\partial_x f\partial_k H
-\partial_k f\partial_x H$ defines the action of the Hamiltonian vector field
$v_H$ generated by $H$ on a function $f\in{\cal U}^{(c)}$.

Given this strongly parallel treatment of the equations of motion for
quantum and (quasi)classical variables, it is natural to couple them by
working with an algebra ${\cal U}^{({\it q-c})}$ of functions of both
commutative elements $x,\ k$ satisfying the Poisson bracket relation
$\{x,k\}=1$ and noncommutative elements $q,\ p$ satisfying the canonical
commutation relation $[q,p]=i$.  All other brackets between these
elements are assumed to vanish.  (Note that $x,\ k,\ q,\ p$ are essentially
generators of the algebra  ${\cal U}^{({\it q-c})}$ and should be
distinguished from the dynamical variables $x(t)$, $k(t)$, etc. which
happen to take those specific values at the initial time.)  Two
general elements $A,\ B$ of
${\cal U}^{({\it q-c})}$, involving both commutative and
noncommutative variables, are canonically conjugate with respect to a
quasiclassical bracket if they satisfy the relation
\begin{equation}
[A,B]_{(q-c)}=[A,B]+i\{A,B\}=i
\end{equation}
The order of quantum variables is preserved when evaluating the Poisson
bracket: thus, if $U,\ V$ are functions of the quantum variables and $f,\
g$ are functions of the quasiclassical variables, one has the quasiclassical
bracket $[fU,gV]_{(q-c)}=fg[U,V]+iUV\{f,g\}$. This quasiclassical bracket
differs from
the definition in Refs. \cite{Ale} and \cite{BoT} and in particular is not
antisymmetric.

Since every element of ${\cal U}^{({\it q-c})}$ is
expressed in terms of $q,\ p,\ x$ and $k$, canonical transformations can be
characterized by their action on these elements.  Elementary
canonical transformations\cite{qct} can be defined which transform the
noncommutative elements $q,\ p$ as quantum variables and the commutative
ones $x,\ k$ as classical variables---that is, without sensing the
mixed character of the quasiclassical bracket.  A large class of (essentially
all useful) canonical transformations are produced by composition of
elementary transformations.  In particular composition produces
transformations which are of mixed character, and one
finds the bracket which is preserved under composition is the
quasiclassical bracket above.  This justifies the choice of this
bracket.  A general canonical transformation is then a map between pairs of
canonically conjugate elements, which preserves
the quasiclassical bracket relations $[A,B]_{(q-c)}=i$.
Note that if one has $[f,g]=0$ or $\{U,V\}=0$ initially, these relations
are not generically preserved by general canonical transformations.

For a Hamiltonian $H(q,p,x,k,t)$,
a variational principle analogous to those above shows that the coordinates
$(q(t),p(t),x(t),k(t))$
satisfy coupled Hamilton-Heisenberg equations of motion, which
at the initial time read
\begin{eqnarray}
\label{HeH}
\dot q= -i[q,H]&,& \quad
\dot p= -i[p,H],  \\
\dot x= \{x,H\}&,& \quad
\dot k= \{k,H\}. \nonumber
\end{eqnarray}
A function $A(q,p,x,k,t)$ of both quasiclassical and quantum variables
satisfies
\begin{equation}
{dA \over dt} = -i[A,H]_{(q-c)}+{\partial A \over \partial t}.
\end{equation}

If $dH/dt=0$, evolution in the ``Heisenberg
picture'' is generated by $e^{v_H^{({\it q-c})}t}$,
where $v_H^{({\it q-c})}$ is the quasiclassical analog of a Hamiltonian
vector field $v_H^{({\it q-c})} f= -i[f,H]_{(q-c)}$.
Since the quasiclassical bracket is not antisymmetric, it is
possible for a time-independent Hamiltonian to have nontrivial evolution
\begin{equation}
dH/dt= -i[H,H]_{(q-c)}\ne 0.
\end{equation}
This is an unusual feature of this proposal, apparently related
to factor ordering of the Hamiltonian as the
coordinates evolve.   Evolution in this case is not yet well
understood\cite{Sal}.

The coupled Hamilton-Heisenberg equations of motion are
invariant under general canonical transformations which mix the quantum and
quasiclassical degrees of freedom, so their solutions can be computed
in any basis.  An S-matrix computed in one basis is equivalent to one
computed in another, and this answers Duff's objection\cite{Duf} that such
transformations would not preserve the S-matrix in a theory with both
classical and quantum variables\cite{comment}.

It is clear from the equations of motion (\ref{HeH}) that the quasiclassical
variables will evolve to depend on the quantum variables $q,\ p$,
and {\it vice versa}.  The latter condition is familiar:  the quasiclassical
variables can be viewed simply as c-number parameters in the Heisenberg
operators $q(t)$ and $p(t)$.  That the quasiclassical variables come to
depend on $q$ and $p$ is the distinguishing feature of this proposal and
leads to state-dependent values for $x(t)$ and $k(t)$. Choose as
initial state the product of a quantum state (possibly a density matrix)
and a pair of definite values for the quasiclassical variables (or a
density matrix describing their positive joint probability distribution).
In the Heisenberg picture, one can compute
expectation values of observables at time $t$ in the usual way---$x$
and $k$ act as multiplication operators on the initial state, taking
the values of the initial quasiclassical positions and momenta (with
their respective probabilities).

Alternatively, suppose that $x(t)=A(q,p,x,k,t)$. The operator dependence
may be interpreted by decomposing the initial quantum state into
eigenstates of $A$, using the initial quasiclassical values for $x$ and
$k$. Then, the probability that $x(t)$ takes a given value $\lambda$ is the
probability that the quantum state is in the eigenstate having the
eigenvalue $\lambda$.

One may well ask whether the joint distribution of values for $(x(t),k(t))$
is necessarily a positive probability distribution. The answer is that it
is not. This should not be unexpected. If evolution leads to
$[x(t),k(t)]\ne 0$, as it may, $x(t)$ and $k(t)$ cannot be simultaneously
measured, and they won't necessarily have a positive joint probability
distribution. A consequence is that the Schr\"odinger representation can be
problematic because the evolved quasiclassical system is not obviously
described by a state (or even a density matrix). This is a topic for
further study.

One might argue that the quasiclassical variables are no longer very
classical because they have lost the
property of simultaneous measurability.  The essential point though is
that this has arisen because of their entanglement with the quantum
variables and not because of intrinsic self-interference.
One expects that a coarse-graining on the scale of the quantum
variables will render the joint distribution positive.  Incidentally,
it is the fact that $x(t)$ and $k(t)$ are limited in their simultaneous
measurability by one's ability to put the quantum system in a simultaneous
eigenstate of each that prevents one from violating the uncertainty
principle by  making measurements of the quasiclassical variables.

An example will illustrate the proposal. Consider the Hamiltonian
$H={1\over 2} kp^2$.  The solutions are easily found to be $p(t)=p$,
$k(t)=k$, $q(t)=  q + k p t$, $x(t) =x +{1\over 2} p^2 t$.
Since $x(t)$ and $k(t)$ commute with $H$, one can work in the
Schr\"odinger picture by decomposing the initial state into eigenfunctions
of $H$:
the wavefunction of the quantum particle is expanded in plane waves and
the initial conditions of the quasiclassical particle are labeled
$|(x',k')\rangle$. The propagator can be formally denoted $e^{-iHt}$
which is understood to give the usual phase factor involving the
eigenvalue of $H$ when it acts on the initial state. It also
evolves the quasiclassical variables
to their final operator values which then act on the quantum
state.  (Generally Schr\"odinger picture evolution is more subtle
than this and requires discussion beyond the scope of this Letter.)
The evolved state of the full system is
\begin{eqnarray}
\Psi&=&e^{-iHt}\int dp'\,f(p') e^{ip'q}|(x',k')\rangle  \\
&=& \int dp'\,f(p') e^{ip'q}e^{-{i\over 2} k' p^{\prime 2} t}
|(x'+{1\over 2} p^{\prime 2}t, k')\rangle. \nonumber
\end{eqnarray}
If the wavepacket for the quantum particle is highly concentrated
around the momenta $p_1'$ and $p_2'$ with amplitudes $\psi_1(q)$
and $\psi_2(q)$ respectively,  then the state of the
full system is approximately
\begin{equation}
\Psi\approx \psi_1(q)|(x'+{1\over 2} p_1^{\prime 2}t, k')\rangle +
\psi_2(q)|(x'+{1\over 2} p_2^{\prime 2}t, k')\rangle.
\end{equation}
The probability that the classical particle is at $x(t)=x'+
{1\over 2} p_1^{\prime 2}t$ at time $t$ is $\int |\psi_1(q)|^2 dq$.

A second example shows three more features of the coupling of
quasiclassical and quantum variables.  Consider the Hamiltonian
\begin{equation}
H={1\over 2} p^2 + akq.
\end{equation}
The solution to the equations of motion are $k(t)=k$, $p(t)=p-
ak t$, $q(t)=q+p t -{1\over 2}ak t^2$ and
\begin{equation}
x(t)=x + a\int_0^t dt\, q(t)= x+ aq t + {1\over 2} ap t^2
-{1\over 6}a^2k t^3.
\end{equation}
The first thing to note is that $x(t)$ does not depend on the
instantaneous eigenvalue of a fixed operator but has an
accumulated dependence over time.  Thus the
operator which determines the distribution of values of $x(t)$
changes dynamically.

Next, there is a feedback term $-a^2 k t^3/6$ present.
This arises because the quasiclassical variables influence the
quantum variables
which in turn act back on the quasiclassical variables.  The
result is a quasiclassical term in the evolution
which would not be present except through the quantum coupling.
It is a second order effect as one expects.

The solution to the Hamilton-Heisenberg equations of motion
are the Heisenberg picture operators.  This means that it
is easy to compute any expectation value at a later time in
terms of expectation values at the initial times.
In particular, the variance of $x(t)$ is
\begin{equation}
\Delta x(t) =  \Delta(q +{1\over 2} p t)  a t.
\end{equation}
This shows that while one can squeeze the initial quantum state
to make the variance in $x(t)$ vanish at a particular later instant,
the variance cannot be held small.

Since the quantum state does not have a definite value, in equations
of motion like $\dot x=q$, the quantum variable
behaves as a  noise term.  This raises the intriguing
possibility of using quasiclassical quantum theory as a way of
solving classical stochastic differential equations.  It is clear
that one is solving a stochastic differential equation by this
approach.  The question of whether one can solve stochastic differential
equations which are of interest is under investigation.

The proposal for a quasiclassical theory of gravity is to solve
the coupled equations
\begin{equation}
G_{\mu\nu}=8\pi T_{\mu \nu},\quad\quad
T^{\mu\nu}\vphantom{T}_{;\nu}= 0,
\end{equation}
where the metric is assumed to be quasiclassical and the matter fields
are quantum.  There are several difficulties to be overcome
in implementing this in practice.  The most obvious is simply
finding any solutions to the equations.  As a first step, a
covariant gauge-invariant perturbative approach\cite{ElB} seems
natural. In the larger picture, there are  questions
about the nature of the Hilbert space for the quantum variables as this
depends on the choice of background in a way that the Hilbert space
in the quantum mechanical context did not depend on the quasiclassical
variables.  This is but one aspect of the problem of time\cite{Kuchar},
involving the definition of quantum theory in curved backgrounds, which
must be addressed.

Acknowledgements.  I would like to thank the participants, especially
R. LaFlamme, of the program of Geometry and Gravity at the Newton Institute
for helpful discussions of this work.


\begin{thebibliography}{99}


\bibitem{Har}  J.B. Hartle, in {\it Gravitation and Quantizations},
Proceedings of the 1992 Les Houches Summer School, vol. LVII, North-Holland,
Amsterdam (1994).

\bibitem{GMH} M. Gell-Mann and J.B. Hartle, Phys. Rev. D {\bf 47}, 3345
(1993).

\bibitem{Ros} Cf. critical remarks by L. Rosenfeld, Nucl. Phys. {\it 40},
353 (1963).

\bibitem{Kib} T.W.B. Kibble, in {\it Quantum Gravity 2}, edited by C.J.
Isham, R. Penrose and D.W. Sciama, (Clarendon Press, Oxford, 1981),  63.


\bibitem{Page}  For a review with comprehensive bibliography, see
D. Page, {\it Proceedings of the 5th Canadian Conference on General
Relativity and Relativistic Astrophysics}, edited by R.B. Mann
and R.G. McLenaghan (World Scientific, Singapore, 1994).

\bibitem{Ale} I.V. Aleksandrov, Z. Naturf.  {\bf 36A}, 902 (1981).

\bibitem{BoT}  W. Boucher and J. Traschen, Phys. Rev. {\bf D37}, 3522
(1988).

\bibitem{Hu}  B.L. Hu and S. Sinha, preprint Univ. Maryland 93-164
(1993); B.L. Hu, preprint Univ. Maryland 94-45, gr-qc/940361 (1994).

\bibitem{Zur}  W. Zurek, Physics Today {\bf 44} (10), 36 (1993);
in {\it Physical Origins of Time Asymmetry}, edited by J.J. Halliwell,
J. P\'erez-Mercader, and W.H. Zurek, (Cambridge Univ. Press,
Cambridge, 1994), 175-207.

\bibitem{cmbr} A. Guth and S.Y. Pi, Phys. Rev. Lett. {\bf 49},
1110 (1982); S. Hawking, Phys. Lett. B {\bf 115}, 295 (1982);
A. Starobinskii, Phys. Lett. B {\bf 117}, 175 (1982).

\bibitem{qct} A. Anderson, Ann. Phys. (NY) {\bf 232}, 292 (1994);
Phys. Lett. B {\bf 319}, 157 (1993).

\bibitem{HBJ} M. Born, W. Heisenberg, and P. Jordan, Ztschr. f. Phys.
{\bf 35}, S. 557 (1926); P.A.M. Dirac, Proc. Roy. Soc. {\bf A110}, 561
(1926).

\bibitem{ncact} J. Schwinger, {\it Quantum Kinematics and Dynamics},
(Benjamin: New York, 1970);  A. Anderson, in preparation.

\bibitem{Sal} L.L. Salcedo, private communication.

\bibitem{Duf} M.J. Duff,  in {\it Quantum Gravity 2}, edited by C.J.
Isham, R. Penrose and D.W. Sciama, (Clarendon Press, Oxford, 1981),  81.

\bibitem{comment}  Other aspects of Duff's argument, in particular, the
invariance of renormalization under canonical transformations, need
further attention in field theory.

\bibitem{ElB} G.F.R. Ellis and M. Bruni, Phys. Rev. D{\bf 40}, 1804
(1989).

\bibitem{Kuchar} K. Kuchar in {\it Proceedings of the 4th Canadian
Conference on General Relativity and Astrophysics}, edited by
G. Kunstatter, D. Vincent, and J. Williams (World Scientific,
Singapore, 1992).

\end{thebibliography}
\end{document}